\documentclass[amsmath,amssymb, aps,pra,twocolumn]{revtex4-2}

\usepackage{xcolor}
\usepackage{amssymb}
\usepackage{float}
\usepackage{graphicx}
\usepackage{dcolumn}
\usepackage[colorlinks,linkcolor=blue,anchorcolor=blue,citecolor=blue,urlcolor=blue]{hyperref}
\usepackage{bm}
\usepackage{siunitx}
\DeclareSIUnit\gauss{G}
\DeclareSIUnit\mgauss{mG}
\DeclareSIUnit\gradient{G/cm}
\usepackage[mathlines]{lineno}%

\begin{document}

\preprint{APS/123-QED}

\title{Feshbach spectroscopy of ultracold mixtures of $^{6}{\rm Li}$ and $^{164}{\rm Dy}$ atoms}

\author{Ke Xie$^{1,2}$}
\author{Xi Li$^{2,3}$}
\author{Yu-Yang Zhou$^{1,2}$}
\author{Ji-Hong Luo$^{1,2}$}
\author{Shuai Wang$^{1,2}$}
\author{Yu-Zhao Nie$^{1,2}$}
\author{Hong-Chi Shen$^{1,2}$}
\author{Yu-Ao Chen$^{1,2,3}$}
\author{Xing-Can Yao$^{1,2,3}$}
\author{Jian-Wei Pan$^{1,2,3}$}

\affiliation{$1$Hefei National Research Center for Physical Sciences at the Microscale and School of Physical Sciences, University of Science and Technology of China, Hefei 230026, China}
\affiliation{$2$Shanghai Research Center for Quantum Sciences and CAS Center for Excellence in Quantum Information and Quantum Physics, University of Science and Technology of China, Shanghai 201315, China}
\affiliation{$3$Hefei National Laboratory, University of Science and Technology of China, Hefei 230088, China}


\begin{abstract}
We report on the observation of Feshbach resonances in ultracold $^6\mathrm{Li}$-$^{164}\mathrm{Dy}$ mixtures, where $^6\mathrm{Li}$ atoms are respectively prepared in their three lowest spin states, and $^{164}\mathrm{Dy}$ atoms are prepared in their lowest energy state. We observe 21 interspecies scattering resonances over a magnetic field range from 0 to \SI{702}{\gauss} using atom loss spectroscopy, three of which exhibit relatively broad widths. These broad resonances provide precise control over the interspecies interaction strength, enabling the study of strongly interacting effects in $^6\mathrm{Li}$-$^{164}\mathrm{Dy}$ mixtures. Additionally, we observe a well-isolated interspecies resonance at \SI{700.1}{\gauss}, offering a unique platform to explore novel impurity physics, where heavy dipolar $^{164}\mathrm{Dy}$ atoms are immersed in a strongly interacting Fermi superfluid of $^6\mathrm{Li}$ atoms.

\end{abstract}

\maketitle

\section{\label{1}INTRODUCTION}

Ultracold mixtures of different atomic species offer an ideal platform for exploring a broad spectrum of physics in both few- and many-body regimes, including Efimov states~\cite{efimov1970energy,kraemer2006evidence,barontini2009observation,bloom2013tests,pires2014observation,tung2014geometric}, polar molecules~\cite{ni2008high,ni2010dipolar,takekoshi2014ultracold,molony2014creation,guo2016creation,park2017second,schindewolf2022evaporation,yang2022creation,bigagli2024observation}, and impurity physics~\cite{kondo1964resistance,massignan2014polarons,hu2016bose,schmidt2018universal}. Additionally, heteronuclear mixtures can introduce mass and population asymmetries, along with tunable interspecies interactions, significantly enhancing the complexity and versatility of quantum systems. These advantages pave the way for the realization of quantum droplets~\cite{cabrera2018quantum}, the investigation of fermion-mediated interactions~\cite{desalvo2019observation}, and the exploration of unconventional fermion pairing mechanisms beyond Bardeen-Cooper-Schrieffer~(BCS) theory~\cite{liu2023quartet}.

In previous studies, heteronuclear mixtures were typically produced by combining alkali metals~\cite{modugno2001bose,hadzibabic2002two,silber2005quantum,taglieber2008quantum,wille2008exploring,yao2016observation,ferrier2014mixture} or alkali and alkaline-earth elements~\cite{hara2011quantum,pasquiou2013quantum,roy2017two}, where the main interspecies interactions were contact interactions. Recently, the realization of quantum mixtures involving highly magnetic elements \cite{ravensbergen2018production,trautmann2018dipolar,ciamei2022double,schafer2023realization} has introduced anisotropic long-range interactions into the system, offering a platform for realizing exotic phases such as the spin rotons, supersolids~\cite{kirkby2023spin}, and dipolar polarons~\cite{kain2014polarons}. The increasing role of spin-spin coupling and multiple-valence electronic configurations in these mixtures makes them particularly suitable for creating strongly interacting systems with high partial wave interactions~\cite{gurarie2005quantum}. Additionally, the potential to create ultracold paramagnetic molecules with substantial magnetic moments in these systems~\cite{finelli2024ultracold} offers promising opportunities for precision measurements and quantum computation~\cite{karra2016prospects}.

\begin{figure}[h] 
	\centering 
		\includegraphics[width=0.45\textwidth]{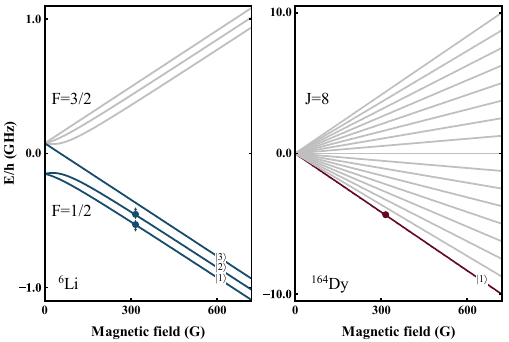} 
		\caption{Magnetic-field dependence of the atomic ground-state energies of $^6{\rm Li}$ and $^{164}{\rm Dy}$ in different Zeeman levels. The spin channels of interest are marked in blue (lithium) and red (dysprosium), i.e., $\text{Li}|i\rangle\text{-}\text{Dy}|1\rangle(i=1,2,3)$. The blue and red spheres denote the initial three-spin mixture of $\text{Li}|1\rangle$-$\text{Li}|2\rangle$-$\text{Dy}|1\rangle$ at \SI{315}{\gauss}.} 
		\label{Fig.0} 
	\end{figure}

In this article, we report on the realization of an ultracold mixture of $^6{\rm Li}$ and $^{164}{\rm Dy}$ atoms. We perform Feshbach spectroscopy to experimentally characterize their scattering properties, which are crucial for investigating the few-body and many-body physics within this system. We identify 21 Feshbach resonances in three spin combinations of $^6{\rm Li}\text{-}^{164}{\rm Dy}$ mixtures (see Fig.~\ref{Fig.0} for the interested spin channels), primarily distributed above \SI{200}{\gauss}, with the resonances spaced relatively sparsely, contrasting with the chaotic and dense resonance spectrum typically observed at lower magnetic fields in dysprosium atoms~\cite{baumann2014observation}. Notably, three relatively broad resonances are observed, making them promising candidates for achieving strongly interacting Bose-Fermi mixtures and ground-state paramagnetic molecules. Additionally, the large mass ratio between $^{6}{\rm Li}$ and $^{164}{\rm Dy}$ reduces the universal geometric scaling factor of Efimov resonances~\cite{naidon2017efimov}, making this system particularly suited for exploring Efimov physics and Borromean trimers~\cite{cui2014universal,zhao2024effects}. Furthermore, we observe a well-isolated resonance between the lowest Zeeman sublevel of $^{164}{\rm Dy}$ and the third-lowest Zeeman sublevel of $^6{\rm Li}$ at \SI{700.1}{\gauss}. At this magnetic field and low temperature, a balanced spin mixture of $^6{\rm Li}$ atoms, with one of the components in the third-lowest Zeeman sublevel, can form a strongly interacting Fermi superfluid~\cite{li2022second,li2024observation}. Therefore, our setup offers a remarkable opportunity to study dipolar impurities immersed in a Fermi superfluid, providing insights into unconventional pairing mediated by magnetic correlations~\cite{scalapino1999superconductivity}.

The article is organized as follows: In Sec.~\ref{2}, we describe the procedure for preparing an ultracold mixture of $^6{\rm Li}$ and $^{164}{\rm Dy}$ atoms and performing atom loss spectroscopy. We then discuss the experimental results, with a particular focus on the three relatively broad resonances. In Sec.~\ref{3}, we conduct a detailed study of the narrow $^6{\rm Li}$-$^{164}{\rm Dy}$ resonance at \SI{700.1}{\gauss}, where the scattering lengths are determined by measuring the thermalization rates between the two species. Finally, in Sec.~\ref{4}, we summarize the findings and provide an outlook on future work.

\begin{figure*}[htbp] 
	
	\centering
	\includegraphics[width=0.8\textwidth]{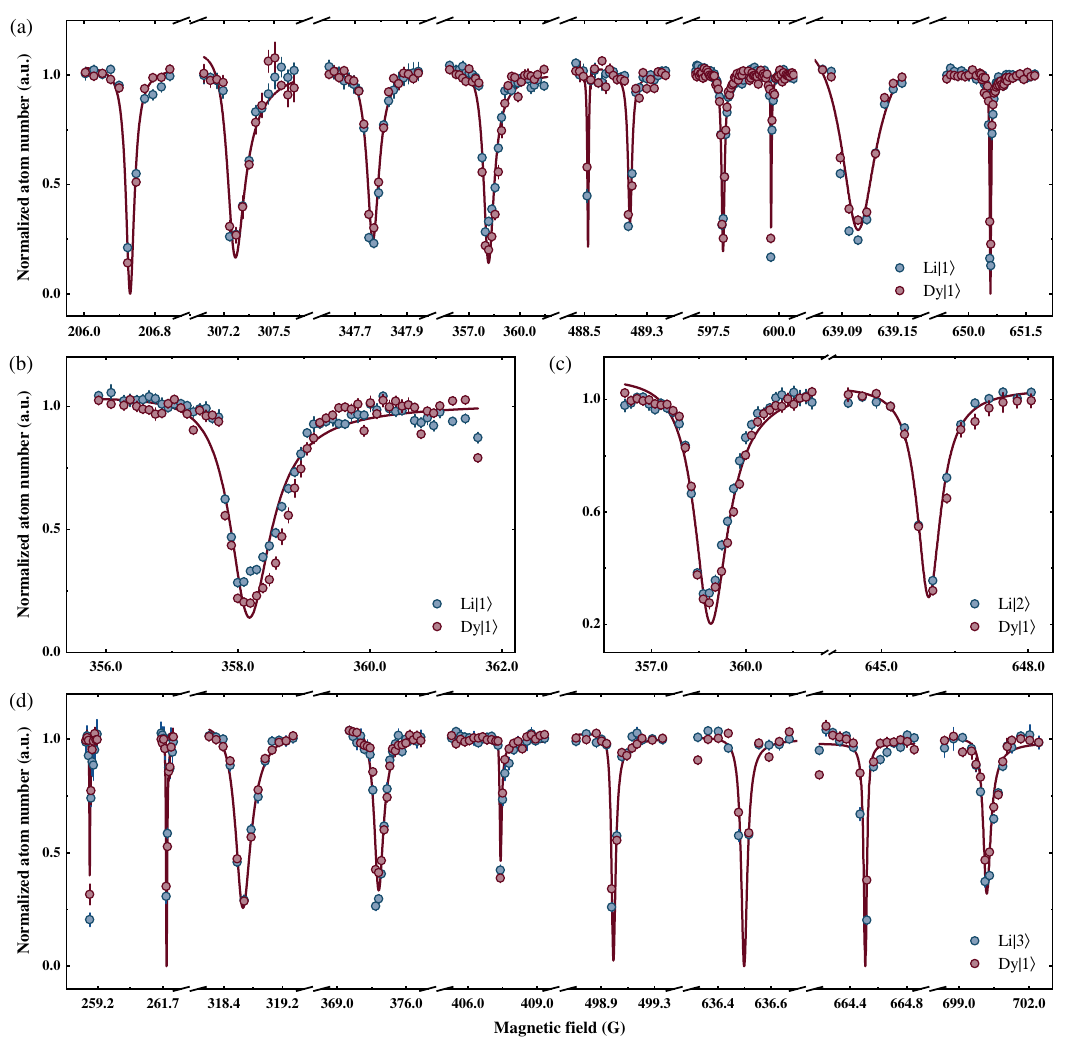} 
	\caption{Normalized remaining atom numbers of $^{164}{\rm Dy}$ (red circles) and $^6{\rm Li}$ (blue circles) as a function of magnetic field $B$ at approximately \SI{2}{\mu\kelvin}. The red solid line represents a fit to the atom loss features of $^{164}{\text{Dy}}$ atoms using Eq.~(\ref{WBF}). (a) Measurements between $\rm Li|1\rangle$ and $\rm Dy|1\rangle$. (b) The relatively broad loss spectrum between $\rm Li|1\rangle$ and $\rm Dy|1\rangle$ at \SI{358.21(3)}{\gauss} with enhanced magnetic field resolution. (c) Measurements between $\rm Li|2\rangle$ and $\rm Dy|1\rangle$. (d) Measurements between $\rm Li|3\rangle$ and $\rm Dy|1\rangle$. Error bars represent the standard error derived from five individual measurements at the same magnetic field.}
	\label{Fig.1}
\end{figure*}

\section{\label{2}Feshbach spectroscopy}

The experimental system consists of three different interspecies spin combinations, noted as $\text{Li}|i\rangle$-$\text{Dy}|1\rangle$ ($i=1,2,3$), as shown in Fig.~\ref{Fig.0}. Here, $^6\text{Li}$ atoms are prepared in one of the three lowest hyperfine sublevels: $\text{Li}|1\rangle=|F=1/2, m_F=1/2\rangle$, $\text{Li}|2\rangle=|F=1/2, m_F=-1/2\rangle$, or $\text{Li}|3\rangle=|F=3/2, m_F=-3/2\rangle$, while $^{164}\text{Dy}$ atoms are polarized in their lowest spin state, $\text{Dy}|1\rangle=|J=8, m_J=-8\rangle$. To prepare these ultracold mixtures, we confine laser-cooled $^6\text{Li}$ and $^{164}\text{Dy}$ atoms in a crossed-beam dipole trap (CDT), formed by two \SI{1064}{\nano\meter} laser beams in the horizontal plane, with a tunable trap geometry by adjusting the aspect ratio of the beam waists. At this stage, $1.50(4) \times 10^5$ $^6\text{Li}$ atoms are equally populated in the $\text{Li}|1\rangle$ and $\text{Li}|2\rangle$ states, while $2.05(6)\times 10^5$ $^{164}\text{Dy}$ atoms are polarized in the $\text{Dy}|1\rangle$ state, forming a three-spin system: $\text{Li}|1\rangle\text{-}\text{Li}|2\rangle\text{-}\text{Dy}|1\rangle$ (see Fig.~\ref{Fig.0}).

We then perform dual-species evaporative cooling by ramping down the power of the CDT at \SI{315}{\gauss} over 5 seconds. The final trap depths are $U_{\rm CDT}^{\rm Li}/k_B= \SI{17.0(1)}{\mu\kelvin}$ for $^6\text{Li}$ and $U_{\rm CDT}^{\rm Dy}/k_B= \SI{11.6(1)}{\mu\kelvin}$ for $^{164}\text{Dy}$. Using parametric heating, we measure the trap frequencies for $^6\text{Li}$ and $^{164}\text{Dy}$, finding  $(\omega_{x},\omega_{y},\omega_{z})_{\rm Li}=2\pi \times (858.3(5), 643(2), 1072(2)) \,\text{Hz}$ and $(\omega_{x},\omega_{y},\omega_{z})_{\rm Dy}=2\pi \times (134.4(3), 101.8(3), 168.6(3))\,\text{Hz}$, respectively. From the measured trap frequency ratio between $^{164}\text{Dy}$ and $^{6}\text{Li}$, we determine the scalar and tensor components of the dynamical polarizability of $^{164}\text{Dy}$ in the ground state at \SI {1064}{\nano\meter}, as $\tilde{\alpha}_S = 183.7(1.3)\,$a.u and $\tilde{\alpha}_T = -3.3(1.6)\,$a.u., respectively, which align well with the experimental results obtained in a $^{40}\text{K}\text{-}^{164}\text{Dy}$ mixture~\cite{ravensbergen2018accurate}.

Next, we ramp the magnetic field to \SI{589}{\gauss} to prepare the desired two-spin combinations. An ultracold $\text{Li}|2\rangle\text{-}\text{Dy}|1\rangle$ mixture is created by removing the $\text{Li}|1\rangle$ atoms with a \SI{20}{\mu}s resonant optical pulse. For the other two-spin combinations, we first apply an optimized hyperbolic secant RF pulse to transfer nearly all the atoms from the $\text{Li}|2\rangle$ state to the $\text{Li}|3\rangle$ state. Then, we use resonant optical pulses to eject either the $\text{Li}|3\rangle$ or $\text{Li}|1\rangle$ atoms from the trap, creating the $\text{Li}|1\rangle\text{-}\text{Dy}|1\rangle$ or $\text{Li}|3\rangle\text{-}\text{Dy}|1\rangle$ mixtures, respectively. To facilitate the identification of interspecies resonances, we prepare ultracold mixtures with balanced population for all three spin combinations, each containing $5.2(1) \times 10^4$ atoms of each species at a temperature of \SI{2.08(5)}{\mu\kelvin}. This is achieved by carefully adjusting the loading times of the $^6\text{Li}$ and $^{164}\text{Dy}$ magneto-optical traps (MOTs), respectively.

To perform Feshbach spectroscopy, the magnetic field is initially ramped from \SI{589}{\gauss} to a target value within \SI{20}{ms}. The mixture is then held for \SI{2}{\second} while the magnetic field is scanned within a $\pm \SI{1.2}{\gauss}$ range around this target value. After each scan, the remaining atom numbers of $^6\text{Li}$ and $^{164}\text{Dy}$ are measured. The target value is then incremented by \SI{2.4}{\gauss} for each subsequent measurement, covering a total range from 0 to \SI{702}{\gauss}. This approach enables the rapid identification of potential magnetic field regions where interspecies resonances might occur. Once these regions are located, we conduct a precise atom loss spectroscopy near the potential resonances. To minimize systematic errors, the magnetic field is first ramped to a value slightly below the resonance over \SI{5}{ms}. The stability of the magnetic field is approximately \SI{1}{mG}. After allowing \SI{50}{ms} for equilibration, the magnetic field is quickly adjusted to the desired value within \SI{250}{\mu\second} and held for a duration between \SI{50}{ms} and \SI{500}{ms}. This duration is carefully selected to ensure that the maximum atom loss when scanning across a resonance does not exceed 80\%, thereby allowing for a clear observation of the atom loss peak and avoiding broadening of the spectrum. The measurement intervals, which range from \SI{10}{\mgauss} to \SI{300}{\mgauss}, are chosen based on the widths of the observed loss features. Finally, the magnetic field is ramped slightly above the resonance to measure the remaining atom numbers of both species using absorption imaging.

\begin{table}[h]
	\caption{\label{table1}
	Interspecies Feshbach resonances observed in the magnetic field range from 0 to \SI{702}{\gauss}. The resonance positions $B_0$ and widths $\Delta B$ are determined by fitting the $^{164}{\text{Dy}}$ atom loss features using the Breit-Wigner-Fano model. The error bars represent one standard deviation obtained from curve fitting
	}
	\begin{ruledtabular}
		\begin{tabular}{lcr}
			\textrm{Channel}&
			\textrm{$B_0$ (G)}&
			\textrm{$\Delta B$ (G)}\\
			\colrule
			$\rm Li|1\rangle$-$\rm Dy|1\rangle$& $206.519(7)$ & $0.12(1)$\\ 
			& $307.26(1)$ & $0.13(2)$\\
			& $347.767(2)$& $0.057(4)$\\
			& $358.13(3)$ & $0.74(7)$\\
			& $488.54(6)$ & $0.02(15)$\\
			& $489.081(7)$& $0.06(3)$ \\
			& $597.847(2)$ & $0.139(4)$\\
			& $599.688(7)$ & $0.044(4)$\\
			& $639.106(3)$ & $0.045(8)$\\
			& $650.574(2)$ & $0.043(6)$\\
			$\rm Li|2\rangle$-$\rm Dy|1\rangle$& $358.83(4)$ & $1.27(9)$\\
			& $645.98(2)$ & $0.60(4)$\\
			$\rm Li|3\rangle$-$\rm Dy|1\rangle$  &
			$258.89(3)$ & $0.03(1)$\\
			 &$261.837(2)$ & $0.051(3)$\\
			
			& $318.648(7)$ & $0.26(2)$\\
			& $373.16(5)$ & $1.17(13)$\\
			& $407.41(2)$ & $0.10(2)$\\
			& $498.989(3)$ & $0.04(1)$\\
			& $636.498(5)$ & $0.03(5)$\\
			& $664.51(3)$ & $0.03(9)$\\
			& $700.18(2)$ & $0.45(5)$\\
		\end{tabular}
	\end{ruledtabular}
\end{table}

As shown in Fig.~\ref{Fig.1}, we observe a total of 21 distinct loss features across the three different spin combinations. Specifically, 10 loss features are identified for the $\text{Li}|1\rangle\text{-}\text{Dy}|1\rangle$ combination, 2 for $\text{Li}|2\rangle\text{-}\text{Dy}|1\rangle$, and 9 for $\text{Li}|3\rangle\text{-}\text{Dy}|1\rangle$. We attribute these loss features to magnetic Feshbach resonances between $^6\text{Li}$ and $^{164}\text{Dy}$ atoms.  Given the asymmetric line shapes of these loss features, we empirically employ the Breit-Wigner-Fano model~\cite{brown2001origin} to fit the data of $^{164}\text{Dy}$ atoms and estimate the positions and widths of the interspecies Feshbach resonances:

\begin{equation}
N_{\text{Dy}}(B) = A \frac{\left( q + (B - B_0)/\Gamma \right)^2}{1 + \left( (B - B_0)/\Gamma \right)^2} + C.
\label{WBF}
\end{equation}
where $N_{\text{Dy}}$ is the remaining atom number of Dy, $B$ is the magnetic field, $B_0$ is the resonance position, $ \Gamma$ is the resonance width, $q$ is the Fano asymmetry parameter, $A$ is the amplitude, and $C$ is the offset.

This approach allows us to determine both the resonance position ($B_0$) and the width ($\Gamma=\Delta B$) for each resonance. The results of these fits are detailed in Table \ref{table1}. 
	
Furthermore, in each spin combination, we identify a relatively broad resonance. Specifically, for $\text{Li}|1\rangle\text{-}\text{Dy}|1\rangle$, the resonance occurs at $B_0=$\SI{358.13(3)}{\gauss} with $\Delta B=$\SI{0.74(7)}{\gauss}; for $\text{Li}|2\rangle\text{-}\text{Dy}|1\rangle$, it is located at $B_0=$\SI{358.83(4)}{\gauss} with $\Delta B=$\SI{1.27(9)}{\gauss}; and for $\text{Li}|3\rangle\text{-}\text{Dy}|1\rangle$, the resonance appears at $B_0=$\SI{373.16(5)}{\gauss} with $\Delta B=$\SI{1.17(13)}{\gauss}. These relatively broad resonances allow for precise tuning of interspecies interactions across a substantial magnetic field range, potentially providing access to a regime where many-body effects are less dependent on momentum. This tunability facilitates comprehensive experimental investigations into a wide range of many-body phenomena in strongly interacting $^6{\text{Li}}$-$^{164}{\text{Dy}}$ mixtures. Notably, the loss features observed for $\text{Li}|1\rangle$-$\text{Dy}|1\rangle$ at \SI{358.13(3)}{\gauss} and $\text{Li}|2\rangle$-$\text{Dy}|1\rangle$ at \SI{358.83(4)}{\gauss}, as shown in Fig.~{\ref{Fig.1}}(b) and (c), exhibit a significant overlap in their corresponding magnetic field ranges. These overlapping resonances allow for the realization of a Fermi sea composed of $\text{Li}|1\rangle$ and $\text{Li}|2\rangle$ atoms, which strongly interacts with $\text{Dy}|1\rangle$ atoms. This system provides an ideal platform for studying Kondo-correlated states~\cite{bauer2013realizing}. Furthermore, overlapping scattering resonances can lead to complex three-body collision dynamics, such as atom-molecule interference effects, which in turn enhances the possibility of observing multiple Efimov features~\cite{d2009ultracold}.

\section{\label{3}Interspecies resonance in the strongly interacting regime of $^6\text{Li}$}

\begin{figure}[h] 
	\centering 

		\includegraphics[width=0.4\textwidth]{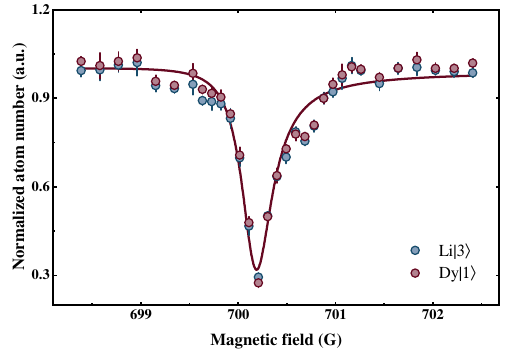} 
		\caption{Normalized remaining atom numbers of $\rm Li|3\rangle$ (blue circles) and $\rm Dy|1\rangle$ (red circles) as a function of magnetic field near the resonance located at \SI{700.18(2)}{\gauss}. Each data point represents the average of five independent measurements. The solid red line shows the result of a Breit-Wigner-Fano fit. Error bars indicate the standard error.} 
		\label{Fig.main2} 
	\end{figure}

In this section, we focus on a specific Feshbach resonance between $\text{Li}|3\rangle$ and $\text{Dy}|1\rangle$, as shown in Fig.~\ref{Fig.main2}. To better characterize the loss spectrum, the measurement accuracy has been significantly enhanced compared to that in Fig.~\ref{Fig.1}(d), clearly revealing the asymmetric loss profile typical of a narrow resonance. The red solid line in Fig.~\ref{Fig.main2} shows the fit to the experimental data for dysprosium atoms (red dots) using Eq.~\ref{WBF}. This fit yields a resonance position of \SI{700.18(2)}{\gauss} and width of \SI{0.45(5)}{\gauss}, respectively. Additionally, we observe a small kink in the loss spectrum around \SI{700.6}{\gauss}, suggesting the possible presence of an Efimov resonance.

This interspecies Feshbach resonance is particularly compelling because, at a magnetic field of \SI{700.18(2)}{\gauss}, the spin-1/2 Fermi system composed of $\text{Li}|1\rangle$ and $\text{Li}|3\rangle$ enters the strongly interacting BCS regime~\cite{li2024observation}. By introducing $\text{Dy}|1\rangle$ to form a three-spin mixture of $\text{Li}|1\rangle$-$\text{Li}|3\rangle$-$\text{Dy}|1\rangle$ at this field, we can prepare a novel quantum system where heavy impurities are immersed in a strongly interacting Fermi superfluid~\cite{wang2022exact}. This resonance allows for precise tuning of the $\text{Li}|3\rangle$-$\text{Dy}|1\rangle$ interaction strength, potentially leading to the observation of Yu-Shiba-Rusinov (YSR) states~\cite{hu2013universal,wang2022exact} and offering valuable insights into the identification of topological superfluids. Moreover, it provides a unique opportunity to explore how magnetic impurities influence the phase diagram of such a superfluid. To this end, it is crucial to precisely characterize the s-wave scattering length between $\text{Li}|3\rangle$ and $\text{Dy}|1\rangle$ atoms near this resonance. The elastic scattering length near an isolated resonance can be expressed as~\cite{chin2010feshbach}:
\begin{eqnarray}
	a(B)=a_{bg}\left(1-\tfrac{\Delta}{B-B_r}\right),
	\label{Equa.main1}
\end{eqnarray}
where $a_{\text{bg}}$ is the background scattering length, with $\Delta$ and $B_r$ denoting the width and position of the resonance, respectively.  

\begin{figure}[h] 
	\centering 
	\includegraphics[width=0.4\textwidth]{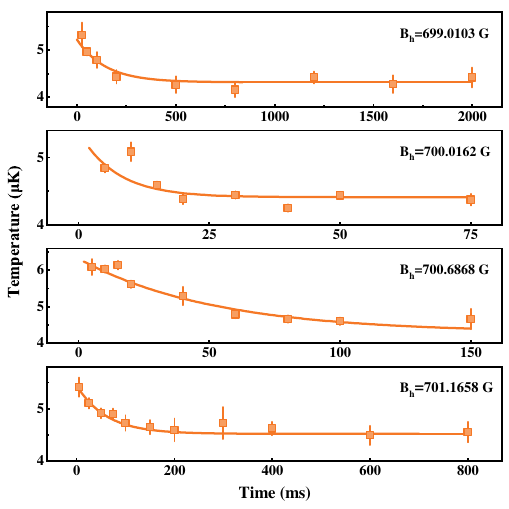} 
	\caption{Interspecies thermalization measurements near the Feshbach resonance. The temperature of $\text{Li}|3\rangle$ atoms is plotted as a function of hold time at target magnetic fields. Each data point, along with its standard error bar, is derived from 5 to 10 independent measurements. The solid orange line represents the fit using an exponential function.} 
	\label{Fig.main3} 
\end{figure}

We determine the elastic scattering length by observing interspecies thermalization at various magnetic fields near the Feshbach resonance. Initially, the heteronuclear mixture has a temperature difference, which equilibrates through interspecies elastic collisions. The elastic scattering cross section is then calculated by tracking the temperature evolution of the atoms over time during the thermalization process~\cite{mosk2001mixture}. To create the initial temperature difference between lithium and dysprosium atoms, we first prepare the desired spin mixture in the CDT. We then apply a \SI{100}{\mu\second} resonant pulse to selectively heat the $\rm Li|3\rangle$ atoms to approximately 
\SI{5.5}{\mu\kelvin}, while the temperature of the $\rm Dy|1\rangle$ atoms remains constant at around \SI{2}{\mu}K. Next, we rapidly ramp the magnetic field to the target value $B_h$ within \SI{5}{\milli\second}, a duration short enough to prevent significant thermalization before measurements. To minimize parallel evaporation effects, we increase the CDT laser power, resulting in final trap depths of $U_{\rm CDT}^{\rm Li}/k_B=\SI{37.2(3)}{\mu\kelvin}$ and $U_{\rm CDT}^{\rm Dy}/k_B=\SI{25.5(2)}{\mu\kelvin}$. The corresponding trap frequencies are $(\omega_{x}, \omega_{y}, \omega_{z})_{\rm Li}=2\pi \times (437.0(3), 547(3), 1370(3))\,$Hz and $(\omega_{x}, \omega_{y}, \omega_{z})_{\rm Dy}= 2\pi \times (69.0(2), 87.0(2), 217.0(5))\,$Hz, respectively. Finally, we hold the mixture for various durations and monitor the temperature evolution of both species. During this hold time, the temperature of the $\rm Li|3\rangle$ atoms decreases towards equilibrium due to elastic collisions with the $\rm Dy|1\rangle$ atoms.

\begin{figure}[htbp] 
	\includegraphics[width=0.45\textwidth]{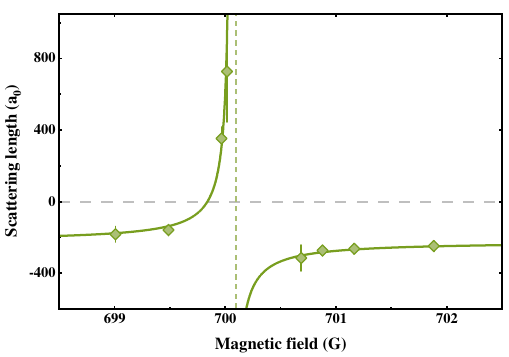} 
	\caption{Interspecies scattering length near the Feshbach resonance. The green solid line represents a fit using Eq.~(\ref{Equa.main1}), while the green dashed line indicates the resonance position $B_\text{r}$. The uncertainty in the scattering length represents one standard deviation, propagated from the uncertainty of the elastic scattering cross-section, which includes uncertainties of temperature, atom number, trap frequencies, and fitting error of the thermalization time.} 
	\label{Fig.main4} 
\end{figure}

Fig.~\ref{Fig.main3} presents the measured temperature of $\rm Li|3\rangle$ atoms (orange squares) as a function of hold time at four exemplary magnetic fields near the resonance. The experimental data is fitted with an exponential function, $T = a e^{-t/\tau} + b$, where $a$ represents the scaling factor, $b$ is the offset, and $\tau$ is the thermalization time.  The elastic scattering cross-section $\sigma_{\rm el}$ near the resonance can be determined using the relation ~\cite{ravensbergen2018production,ye2022observation}:
\begin{eqnarray}
\tau^{-1}=\sigma_{\rm el}\frac{\xi}{3}vn.
	\label{Equa.main3}
\end{eqnarray}
where $\xi=4m_{\text{Dy}}m_{\text{Li}}/(m_{\text{Li}}+m_{\text{Dy}})^2$ is the reduced mass, $v = \sqrt{8k_B/\pi(T_{\rm Li}/m_{\rm Li}+T_{\rm Dy}/m_{\rm Dy})}$ is the mean relative velocity, and $n=N_{\text{Dy}}N_{\text{Li}}m_{\text{Dy}}^{3/2}\bar{\omega}_{\text{Li}}^3/[(2\pi k_B)(T_{\text{Dy}}+\beta^2T_{\text{Li}})]^{3/2}$ is the overlap density integral with $\beta = \sqrt{m_{\text{Li}} \bar{\omega}^2_{\text{Li}}/m_{\text{Dy}} \bar{\omega}^2_{\text{Dy}}}$. The exponential fit can be regarded as an interpolation to determine the initial slope of the thermalization. Therefore, the other physical quantities in Eq.~(\ref{Equa.main3}) can be substituted by their initial values.  Using the energy-independent formula $\sigma_{\rm el} = 4\pi a^2$, we calculate the elastic scattering lengths as a function of magnetic field, shown as green diamonds in Fig.~\ref{Fig.main4}. By fitting the data with Eq.~(\ref{Equa.main1}), we obtain the values $a_{bg} = \SI{-223(8)}{\textit{a}_0}$, $B_r = \SI{700.10(1)}{\gauss}$, and $\Delta = \SI{-0.22(3)}{\gauss}$. The uncertainties in the parentheses are statistical errors determined through curve fitting, which incorporates the propagated uncertainty in the scattering length. We note that, due to the thermally induced broadening of the loss spectrum at finite temperatures \cite{maier2015emergence}, the value of $\Delta B$= \SI{0.45(5)}{G} obtained from the loss curve fitting is larger than the resonance width $\Delta$.

\section{\label{4}Conclusion and outlook}

In this work, we realize an ultracold $^{6}{\rm Li}$-$^{164}{\rm Dy}$ mixture and observe 21 Feshbach resonances across a broad range of magnetic fields for various interspecies spin combinations. Our measurements provide valuable data for theoretical studies~\cite{tiecke2010asymptotic,gao1998quantum,gao2005multichannel} and enable theorists to construct a full coupled-channel model and optimize scattering coefficients via theory-experiment comparison~\cite{chin2010feshbach,ciamei2022exploring}. This approach allows for a precise characterization of the scattering properties of the $^{6}{\rm Li}$-$^{164}{\rm Dy}$ system, which is crucial for future research in molecular physics. Among the observed resonances, three are relatively broad. Further binding energy measurements and theoretical calculations will deepen our understanding of these resonances, facilitate the exploration of both few-body and many-body physics in strongly interacting $^{6}{\rm Li}$-$^{164}{\rm Dy}$ mixtures, and pave the way for the creation of $^{6}{\rm Li}$-$^{164}{\rm Dy}$ paramagnetic molecules. Additionally, we observe a well-isolated s-wave resonance within the strongly interacting BCS regime of $^{6}{\rm Li}$, providing an ideal platform to study magnetic impurities immersed in a Fermi superfluid. Moving forward, we plan to conduct precise measurements of three-body recombination rates to investigate the observed kink in the loss spectrum and explore the potential existence of higher-order Efimov states~\cite{naidon2017efimov}. 

\begin{acknowledgments}
This work is supported by the Innovation Program for Quantum Science and Technology (Grant No. 2021ZD0301900), the National Key R\&D Program of China (Grant No. 2018YFA0306501), NSFC of China (Grant No. 11874340),  the Chinese Academy of Sciences (CAS), the Anhui Initiative in Quantum Information Technologies, and the Shanghai Municipal Science and Technology Major Project (Grant No.2019SHZDZX01).
\end{acknowledgments}

%

\end{document}